\documentstyle[11pt,newpasp,twoside,texdefs, epsf]{article}
\def\edcomment#1{\iffalse\marginpar{\raggedright\sl#1\/}\else\relax\fi}
\reversemarginpar

\begin{document}
\title{X-Ray Absorption Associated with High-Velocity UV Absorbers}
 \author{Bassem M. Sabra \& Fred Hamann}
\affil{Dept. of Astronomy, University of Florida, Gainesville, FL 32611}
\author{Joseph C. Shields}
\affil{Dept. of Physics \& Astronomy, Ohio University, Athens, OH 45701}
\author{Ian George}
\affil{LHEA, NASA/GSFC, GreenBelt, MD 20771}
\author{Buell Jannuzi}
\affil{NOAO, 950 North Cherry Ave., Tuscon, AZ 85719}

\begin{abstract}
We present \textit{Chandra} observations of two radio-quiet QSOs, PG~2302+029 
and PG~1254+047. PG2302+029 has an ultra high-velocity UV absorption system 
($\sim -56,000$ km~s$^{-1}$), 
while PG~1254+047 is a Broad Absorption Line (BAL) QSO with detached troughs. 
Both objects are X-ray weak, consistent with the known correlation between 
$\alpha_{\rm ox}$ and the strength of the UV absorption lines. The data 
suggest that there is evidence that both objects are intrinsically 
weak X-ray sources, in addition to 
being heavily X-ray absorbed. The X-ray absorption column densities are 
$N_{\rm H} > 10^{22}$ cm$^{-2}$ for neutral gas and the intrinsic emission 
spectra have $\alpha_{ox} > 2$. The data are fit best by including ionized 
(rather than neutral) absorbers, with column densities 
$N_{\rm H}^{PG2302} > 2.98\times 10^{22}$ cm$^{-2}$ and 
$N_{\rm H}^{PG1254} > 17.3\times 10^{22}$ cm$^{-2}$. The degrees of 
ionization are consistent with the UV lines, as are the total column 
densities if the strongest lines are saturated.
\end{abstract}
\section{Introduction}
Almost all quasi-stellar objects (QSOs) are X-ray sources. The reprocessing 
of X-rays by matter along the line of sight to the QSO 
imprints informative features on the resulting spectrum. 
We are examining the X-ray properties of QSOs with intrinsic UV absorption 
lines. In this proceeding we discuss {\it Chandra} observations of the 
QSOs PG 2302+029 and PG 1254+047. The BAL QSO PG~1254+047 displays detached 
broad absorption lines ($FWHM\sim 10,000$ km s$^{-1}$ centered at 
-20,000~km~s$^{-1}$) in the UV (Hamann 1998). PG~2302+029 is peculiar for its 
ultra-high velocity ($\sim -56,000$ km~s$^{-1}$) UV absorption lines with 
$FWHM \approx 4,000$ km~s$^{-1}$ (Jannuzi et al. 1996). The intrinsic nature 
of the absorber in PG~2302+029 was confirmed recently by line 
variability (Jannuzi et al. 2001). Both objects are characterized by 
being faint X-ray sources. Our aim is to determine 
the properties of the X-ray spectrum, search for any signs of absorption, and 
define the relationship between the UV/X-ray absorbers. 
If the X-ray and UV absorbers are the same, then the -56,000~km~s$^{-1}$ 
velocity shift of the UV absorption lines in PG~2302+029 {\bf is 
potentially} resolvable with ACIS, given sharp features {\bf and} 
adequate signal-to-noise ratio.

The absorption from this type of objects originates in an outflow of matter 
from the central engine of the QSO. Determining the relation between the 
gases producing the X-ray and UV absorption has profound implications on the 
physics of wind formation and acceleration (e.g. Mathur et al. 1995; Murray 
et al. 1995). Standard analysis, which uses the absorption line troughs 
to derive the optical depths and column densities, typically implies that 
the total column densities are about 2 orders of magnitude lower than the 
X-ray absorbing columns. However, the discrepancy could be alleviated if 
the UV lines are more optically thick than they appear, e.g. partial coverage 
fills the line troughs and thus hides larger column densities (Hamann 
et al. 1998). 
\vspace*{-0.1cm}
\section{Observations}
PG 2302+029 and PG 1254+047 were observed by {\it Chandra} using the 
Advanced Imaging Spectrometer (ACIS) on 7 January 2000 and 29 May 2000, 
respectively. The most recent (2 November 2000 and 29 February 2001 
for PG~2302+029 and PG~1254+047, respectively) 
re-processed data released by the 
CXC were used. No filtering for high background or bad aspect times 
was done since we found that the data were relatively free from such 
problems. Data extraction and calibration was performed using version 
1.4 of CIAO. XSPEC was used for rebinning and spectral analysis. 
We created the response matrix and ancillary response files by relying 
on calibration data provided when the chip temperature during observations 
was $-120\deg$. 

We extracted the source counts from circular regions with radii of 5\arcsec. 
Background regions were annuli with radii of 10\arcsec\ to 20\arcsec. We 
obtained a total of $391\pm 21$ counts for PG 2302+029 and $47\pm 8$ counts 
for PG 1254+047. Table 1 gives a list of pertinent information 
about the two objects. 
The specta were binned to have at least 30 counts/bin (10 counts/bin for 
PG 1254+047). The spectral analysis discussed in this contribution 
includes energy bins below 0.5 keV. It has come to our attention that 
these bins may suffer from calibration problems. 
See Sabra \& Hamann (2001) and 
Sabra et al. (2001) for a treatment where energy bins only above 0.5 keV 
are included. 
\begin{table}
\vspace*{-0.9cm}
\begin{center}
%\hspace*{3cm} 
\caption{PG~2302+029 and PG~1254+047}
\vspace*{-0.3cm}
\begin{tabular}{cccccccc}
\tableline
\tableline
Object & Obs. Date & Exp. (ksec) & $^1$N$_{\rm H}^{\rm Gal.}$ (cm$^{-2}$)  & $^2$z$_{\rm em}$ & z$_{\rm abs}$ & B (mag) \\
PG~2302 & 2000-01-07 &48 & $5\times 10^{20}$& 1.044 &0.695 & 16.30 \\
PG~1254 & 2000-05-29 &36 & $2\times 10^{20}$& 1.024 &0.870 & 16.15 \\
\tableline
\end{tabular}
\end{center}
\vspace*{-0.45cm}
$^1$ Lockmann \& Savage (1994). $^2$ NED.\\
\end{table}
\vspace*{-1.0cm}
\section{Data Analysis and Results}
Our procedure for spectral fitting is the following. We 
start by fitting a power law continuum ($\phi_E  \propto E^{-\Gamma}$) 
absorbed only by the appropriate Galactic column density 
(Lockman \& Savage 1994). Both the normalization 
of this continuum and its X-ray photon index, $\Gamma$, are left as free 
parameters. The results are shown in Figure 1. The slopes are rather 
flat for QSOs, where usually $1.3 < \Gamma < 2.3$ (e.g. Laor et a. 1997; 
Reeves et al. 1997). The X-ray fluxes are low, leading to the steep 
$\alpha_{\rm ox}=2.1, 2.5$ for PG~2302+029 and PG~1254+047, respectively. 
These values are consistent with the correlation  between $\alpha_{\rm ox}$ 
and the equivalent width of {\sc C iv} $\lambda\lambda 1548,1550$ shown in 
Figure 2, which was adapted from Brandt et al. (2000). The correlation is 
indicative of intrinsic absorption: a large equivalent width results from 
an absorber that, in turn, is accompanied by an X-ray absorber thus 
steepening 
$\alpha_{ox}$. 

\begin{figure}
\plottwo{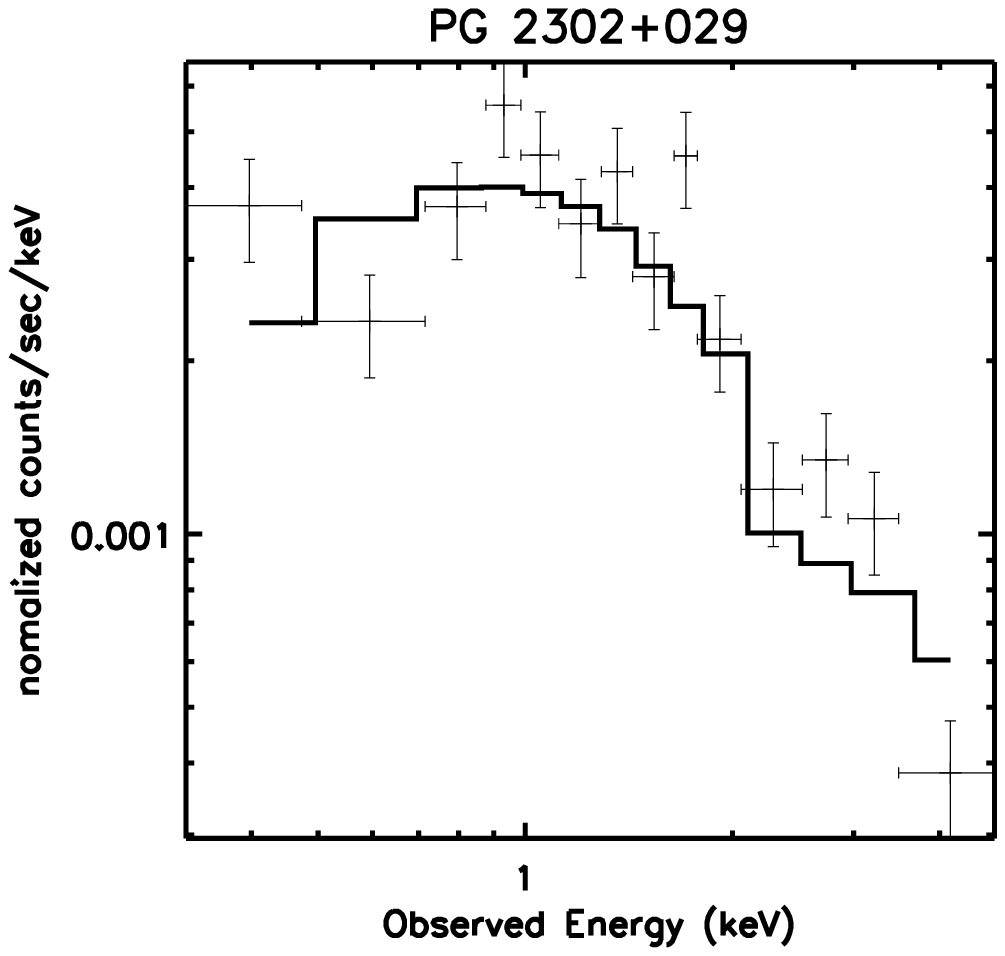}{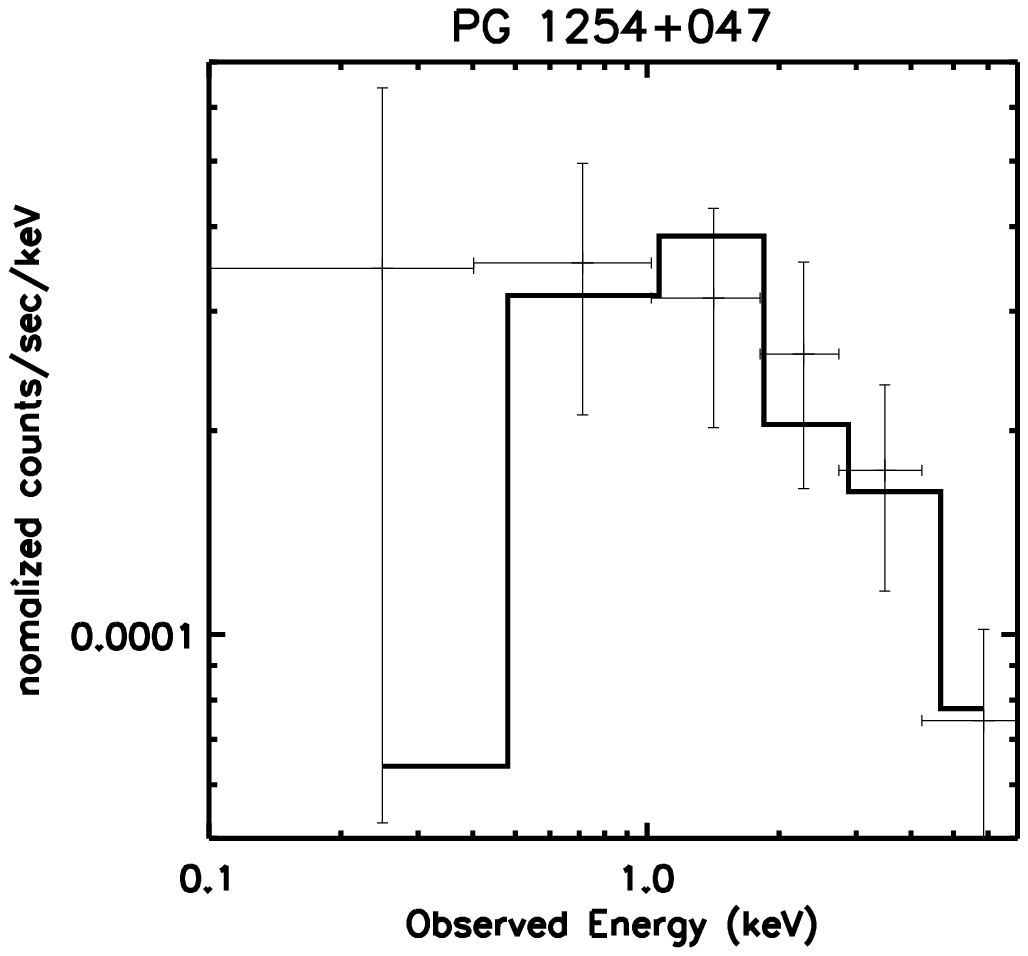}
\caption{Galactic Absorption only, free normalization and $\Gamma$. 
PG~2303+029: 
$\phi_E = 6.88\times 10^{-6}$ photon~s$^{-1}$~cm$^{-2}$~keV$^{-1}$ at 1~keV, 
$\Gamma=1.071$,  $\chi_\nu^2=2.57$ for 12 d.o.f., 
PG~1254+047: 
$\phi_E = 6.63\times 10^{-7}$ photon~s$^{-1}$~cm$^{-2}$~keV$^{-1}$ 
at 1~keV, $\Gamma=0.3787$,  $\chi_\nu^2=0.45$ for 4 d.o.f. 
}
\end{figure}

\begin{figure}
\plotfiddle{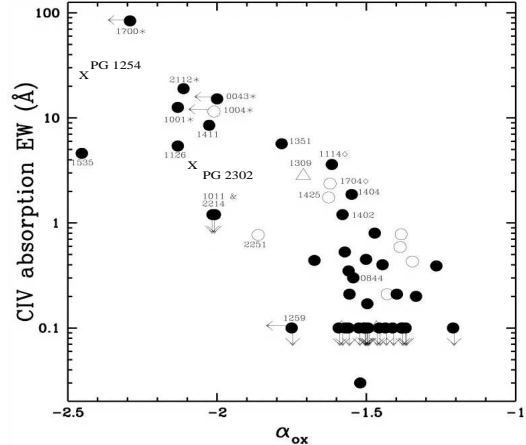}{-20cm}{-90}{50}{40}{-200}{29}
\vspace*{6cm}
\caption{EW {\sc C iv} vs. $\alpha_{ox}$. Solid dots: radio-quiet QSOs, 
open triangles: core-dominated radio-loud QSOs, open circles: lobe-dominated 
radio-loud QSOs. Soft X-ray Weak QSOs are labeled their RAs, while an 
asterisk indicates a BAL QSO and a diamond signifies a warm absorber. We 
overplot our quantities for PG~2302+029 and PG~1254+047 as X's. We have 
recalculated our $\alpha_{ox}$ to be in agreement with the convention in this 
plot, i.e. between 3000 \AA\ and 2 keV, instead of what we have used so 
(2500 \AA\ and 2 keV) (Adapted from Brandt et al. 2000).}
\end{figure}

To test for absorption, we adopt a ``normal'' QSO continuum, specified by  
$\alpha_{\rm ox}=1.6$ and $\Gamma=1.9$ (Laor et al. 1997), attenuated 
through a neutral absorber. The choice of $\alpha_{ox}$ determines the 
normalization of the powerlaw at 2~keV$/(1+z)$, 
$\phi_E(1\ {\rm keV}) \propto f^{obs}_\nu(1\ {\rm keV}) $, which we derive 
in the 
following way. We first calculate the rest-frame $f_\nu(2500\ {\rm \AA})$ 
from the B-magnitude, including the appropriate Galactic dereddening and 
the k-correction (see Green 1996). The rest-frame $f_\nu (2\ {\rm keV})$ and 
$f_\nu (2500\ {\rm \AA})$ are related by  
$\alpha_{ox}=0.384\ log(f_\nu(2500\ {\rm \AA})/f_\nu (2\ {\rm keV}))$. 
Therefore, 
$f^{obs}_\nu(1\ {\rm keV}) \approx f^{rest}_\nu(2\ {\rm keV})/2)$, for $z 
\approx 1$. We experimented by adding neutral absorbers at the 
redshift of the QSO and at the redshift of the UV absorption 
lines. For PG 2302+029, the fits did not favour any particular redshift.  
The redshifts of the absorber and the QSO are too close to be resolved in 
PG~1254+047. In both cases, however, the X-ray absorption at the 
systemic velocity of the QSO leads to higher column densities (by a factor of 
less than 2). We hereafter fix the redshifts of the X-ray absorbers at those 
of the emission lines of the QSOs. The results are shown in Figure 3  
below. The normalizations, based on higher energy channels where absorption 
has little effect, hint at intrinsic X-ray weakness, while the poor fits, 
especially at lower energies, hint at intrinsic absorption. The bad overall 
fit indicates that the observed X-ray weakness cannot be explained by 
absorption alone. 

To study the possibility of both intrinsic X-ray weakness and absorption, 
we remove the constraint that $\alpha_{\rm ox}=1.6$ and hence 
allow the normalization of the power law to vary. The fits improve 
drastically, though not to the extent of giving an acceptable fit 
($\chi_\nu^2 \approx 1$). The intrinsic power 
law flux density at 1~keV with these improved fits decreased by about an 
order of magnitude. We show the results in Figure 4. 
 
The results from the above spectral fitting do not strongly support neutral 
absorption due to the large discrepancy between the data and the models at 
soft energies. Also, we know that there is not a neutral absorber with the 
above quoted column densities because the UV spectra do not contain 
low-ionization metal lines. 
To improve the fits, we experimented with absorption by ionized 
gas, and neutral and ionized partial coverage. We fix the ionization 
parameter U, the ratio of ionizing (above 13.6 eV) flux density to hydrogen 
density. Experiments showed that $logU=0.1$ is consistent with UV data 
(Hamann 1998) and leads to $\chi^2_\nu < 2$. The partial covering fraction, 
$f_{\rm c}$, was fixed at 0.8 for PG~1254+047 because we found that it 
improved the quality of the fits. These values are consistent with absorption 
studies for such objects (e.g., Hamann 1998). Figure 5 displays the outcome. 

The models in Figure 5 allow us to place limits on the column 
densities of the absorbers. We find that for 
PG~2302+029, $2.39 < \frac{N_H}{10^{22}~cm^{-2}} < 3.66$, while for 
PG~1254+047, $6.73 < \frac{N_H}{10^{22}~cm^{-2}} < 41.6$, both at the 
90\% confidence level.\\
\\
\\
\begin{figure}
\vspace*{-0.4cm}
\vbox{
\plottwo{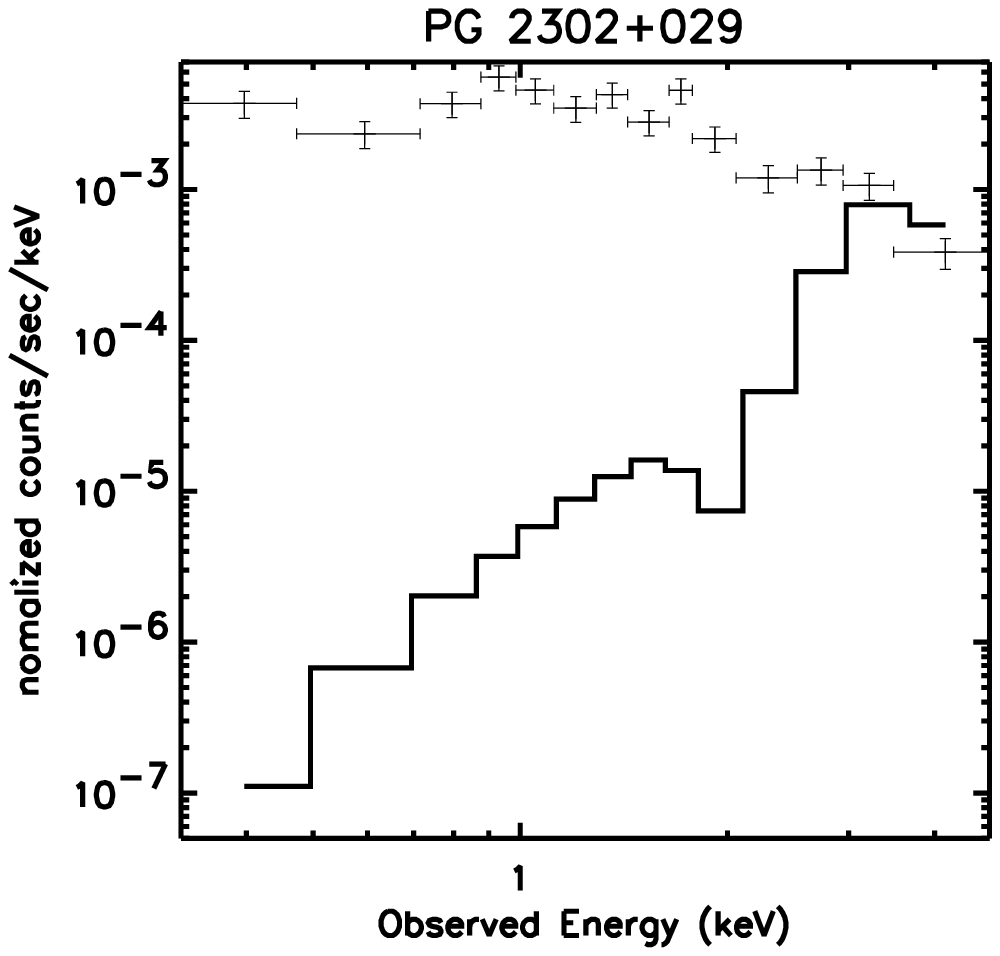}{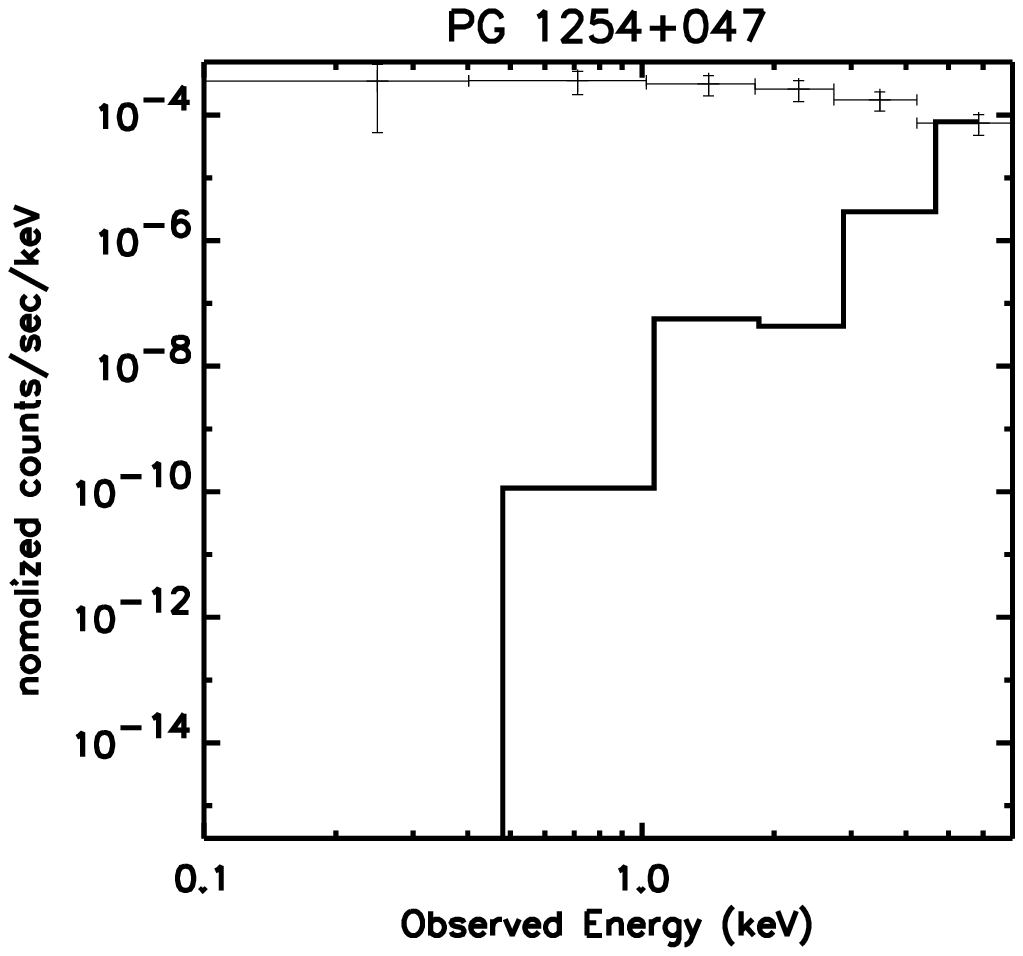}
\caption{Galactic Absorption + Neutral Intrinsic Absorber, 
frozen normalization.
PG~2302+029: 
$N_{\rm H}=149.0\times 10^{22}$ cm$^{-2}$, 
$\chi_\nu^2=23.98$ for 13 d.o.f., 
PG~1254+047:  
$N_{\rm H}=501.1\times 10^{22}$cm$^{-2}$, 
$\chi_\nu^2=6.28$ for 5 d.o.f.}
}
\vbox{
\plottwo{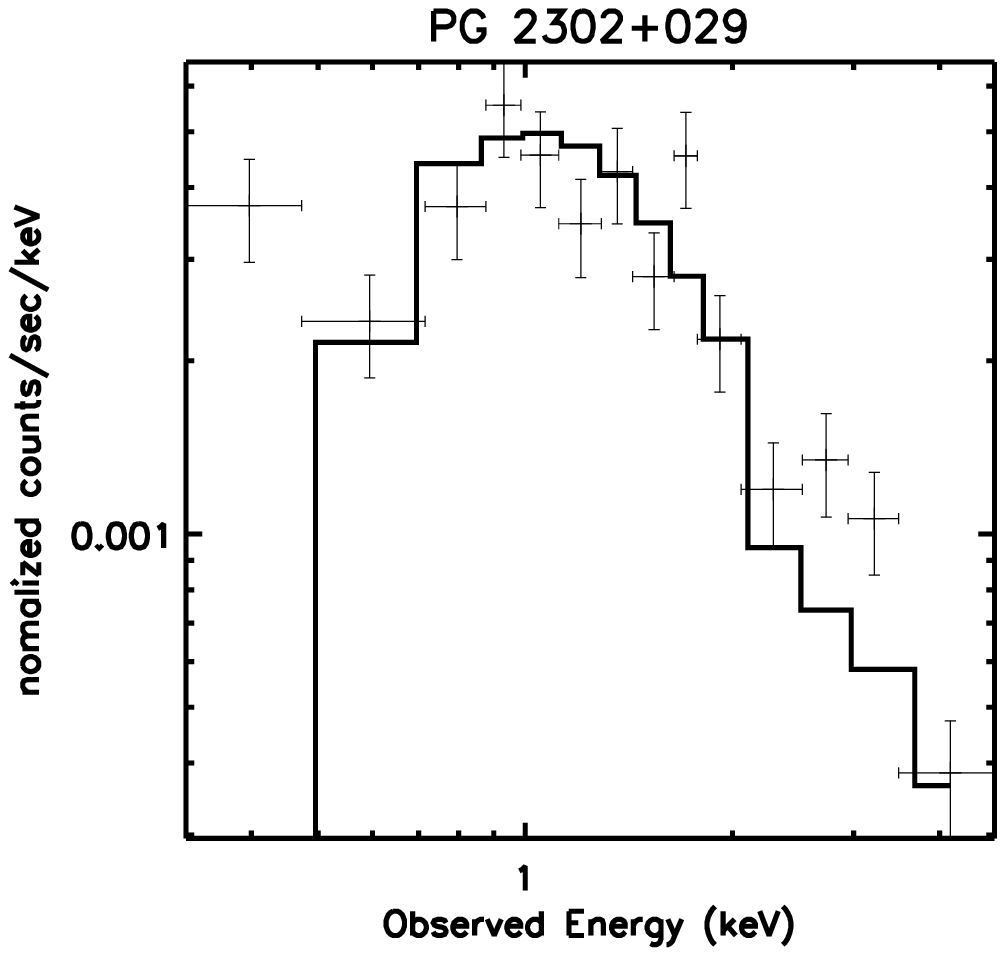}{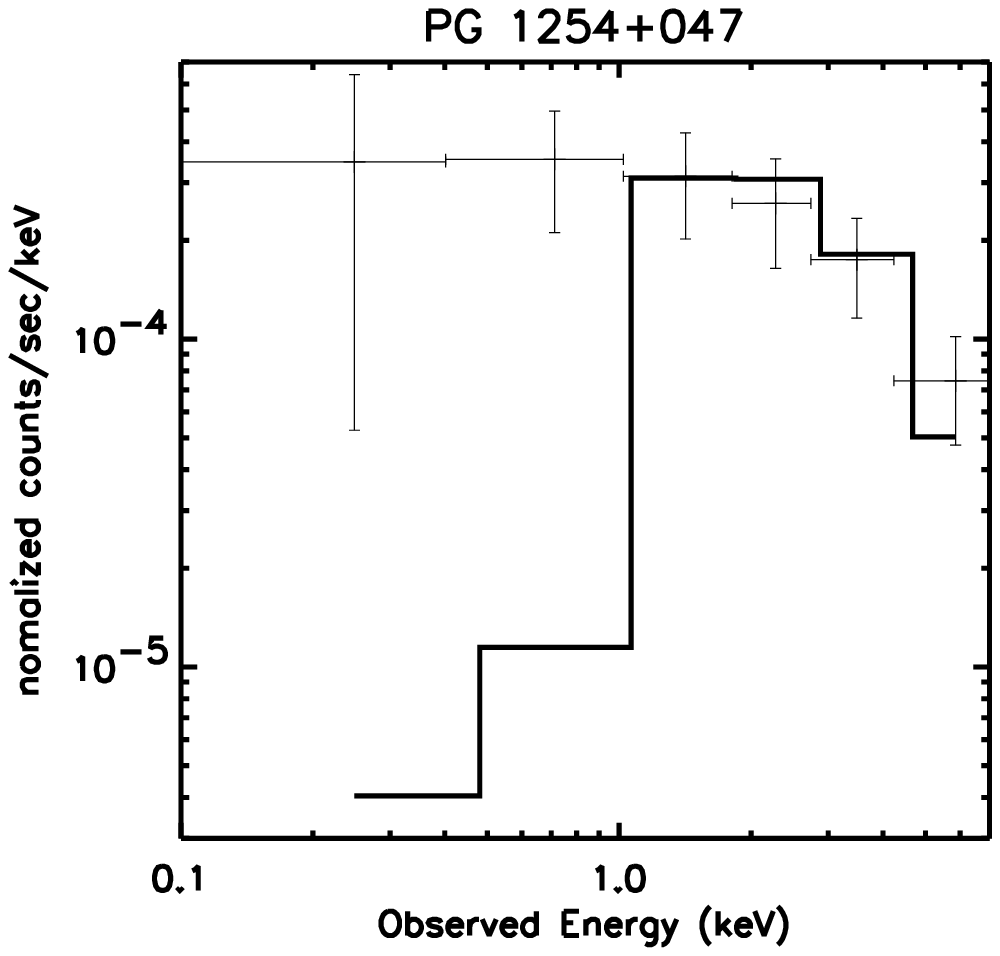}
\caption{Galactic Absorption + Neutral Absorber, free normalization.
PG~2302+029:
$N_H=1.12\times 10^{22}$ cm$^{-2}$, 
$\phi_E = 1.36\times 10^{-5}$ photon~s$^{-1}$~cm$^{-2}$~keV$^{-1}$ at 1~keV, 
$\chi_\nu^2=3.57$ for 12 d.o.f., 
PG~1254+047:
$N_H=10.8\times 10^{22}$ cm$^{-2}$, 
$\phi_E = 5.87\times 10^{-5}$ photon~s$^{-1}$~cm$^{-2}$~keV$^{-1}$ at 1~keV, 
$\chi_\nu^2=2.05$ for 4 d.o.f.}
\vspace*{-2cm}
}
\end{figure}
%\vspace*{-2cm}
\section{Conclusions}
We have presented \textit{Chandra} observations of PG~2302+029, a QSO that 
shows ultra high-velocity UV absorption, and PG~1254+029, a BAL QSO. The 
following points can be made:\\
1- The data suggest that there is evidence the both objects are intrinsically 
X-ray weak, though the X-ray slope is normal. 
The evidence is somewhat 
stronger for PG~2302+029 given the higher number of counts. 
No amount of absorbing column density was able to suppress the X-ray 
flux down to the observed values while at the same time reproduce the overall 
X-ray spectral shape. The intrinsic $\alpha_{\rm ox}$'s are steep:
$\alpha_{\rm ox}^{\rm PG~2302}=2.0$, 
$\alpha_{\rm ox}^{\rm PG~1254}=2.3$.\\
%NOTE TO THE EDITOR: I ``ended'' the figure environment here so that I can
%put Figures 3 and 4 and some part of the text on one page, without leaving 
%extra white spaces.
\noindent 2- There is intrinsic X-ray absorption, most probably ionized with 
$logU=0.1$ and 
$N_{\rm H}^{\rm PG~2302}  \gtsim 2.98 \times 10^{22}$ cm$^{-2}$, 
$N_{\rm H}^{\rm PG~1254}  \gtsim 17.3 \times 10^{22}$ cm$^{-2}$ for 80\% 
partial coverage. The derived column densities are consistent with results 
from the UV data, if the UV lines are very saturated.\\
3- We were not able to determine the redshift of the X-ray absorber 
in PG~2302+029.

\begin{figure}
\plottwo{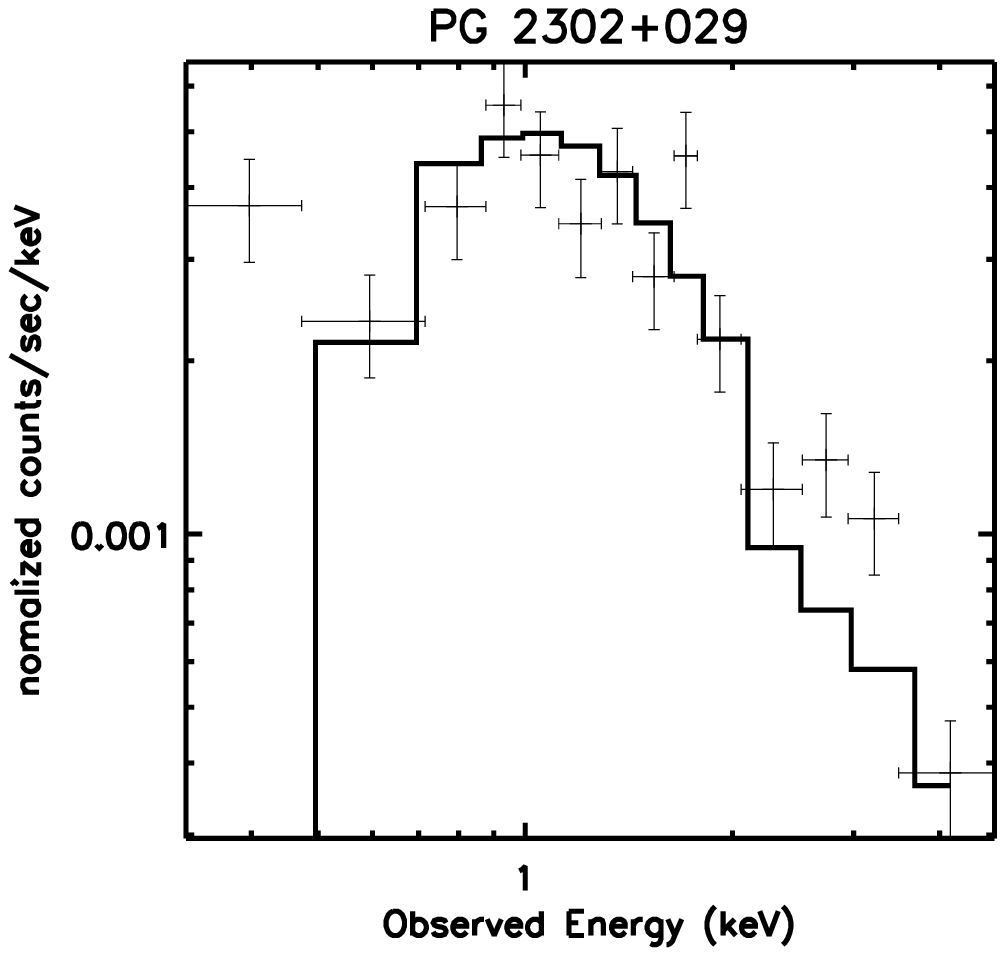}{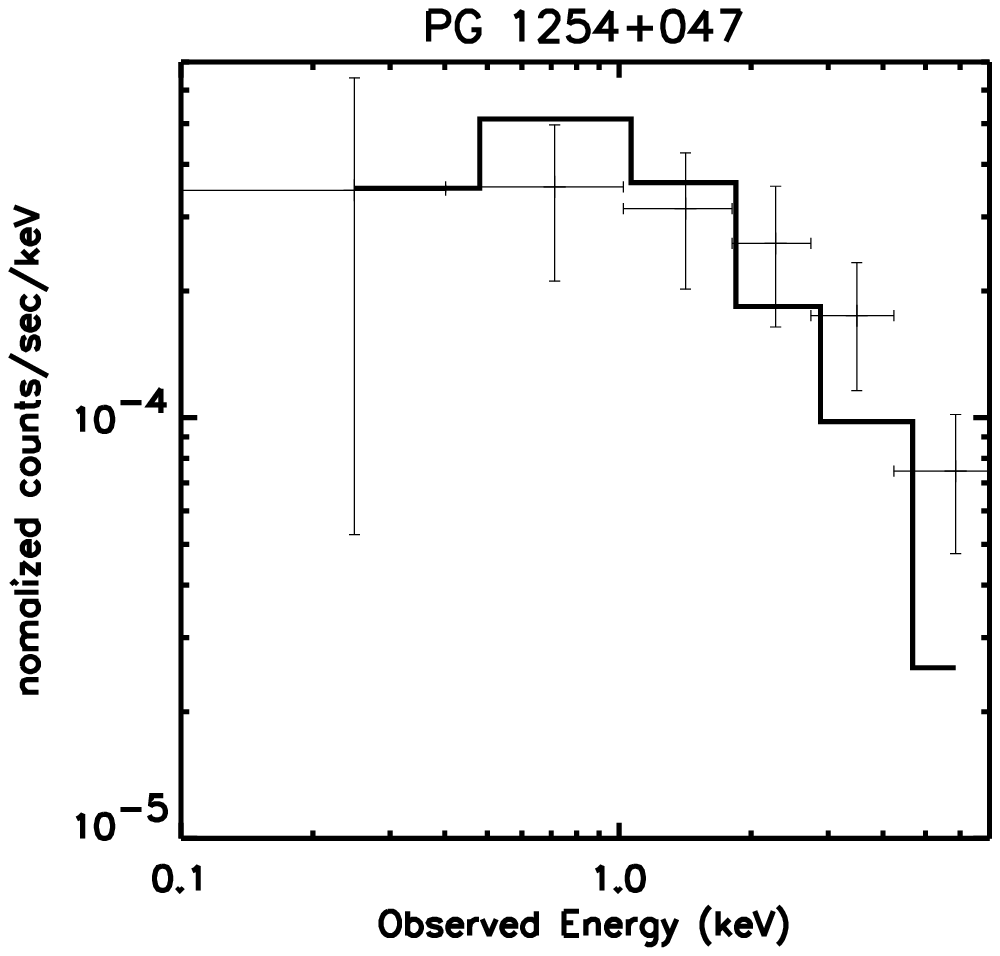}
\caption{Galactic Absorption + Ionized Absorber, free normalization, 
partial coverage (PG~1254+047). 
PG~2302+029: 
$N_{\rm H}=2.98\times 10^{22}$ cm$^{-2}$, $logU=0.1$ (frozen), 
$\phi_E = 1.54\times 10^{-5}$ photon~s$^{-1}$~cm$^{-2}$~keV$^{-1}$ at 1~keV, 
$\chi_\nu^2= 1.66$ for 12 d.o.f., 
PG~1254+047: 
$N_{\rm H}=17.3\times 10^{22}$cm$^{-2}$, $logU=0.1$ (frozen), 
$f_{\rm c}=0.8$ (frozen), 
$\phi_E = 3.05\times 10^{-6}$ photon~s$^{-1}$~cm$^{-2}$~keV$^{-1}$ at 1~keV,  
$\chi_\nu^2=1.76$ for 4 d.of. The normalizations $\phi_E$ correspond to 
intrinsic $alpha_{\rm ox}$=2.0, 2.3, for PG~2302+029 and PG~1254+047, 
respectively.}
\end{figure}
\acknowledgments{
FH and BMS wish to acknowledge support through \textit{Chandra} grants 
GO 0-1123X and GO 0-1157X.}


\vspace*{-0.5cm}
\begin{references}
\reference{} Brandt, W. N., Laor, A., \& Wills, B. J. 2000, \apj, 528, 637
\reference{} Hamann, F. 1998, \apj, 500, 798
\reference{} Jannuzi, B. T., et al. 1996, \apj, 470, L11
\reference{} Jannuzi, B. T., et al. 2001, in preparation
\reference{} Laor, A., et al. 1997, \apj, 477, 93
\reference{} Mathur, S., Elvis, M., \& Singh, K. 1995, \apj, 455, L9
\reference{} Murray, N., Chiang, J., Grossman, J. A., \& Voit, G. M. 1995, 
\apj, 451, 498
\reference{} Sabra, B. M., \& Hamann, F. 2001, \apj\ Letters, to be submitted
\reference{} Sabra, B. M., et al. 2001, \apj\ Letters, to be submitted
\reference{} Reeves, J. N., et al. 1997, \mnras, 292, 468
\end{references}
\end{document}